%% file: aam.tex
\documentclass[final,3p,times]{elsarticle}

\usepackage{booktabs}
\usepackage{multirow}
\usepackage{graphicx}
\usepackage[dvipsnames]{xcolor}
\usepackage{color,soul}
\usepackage[flushleft]{threeparttable}
\usepackage[fleqn,tbtags]{mathtools}
\usepackage{natbib}
\usepackage{hyperref}
\usepackage{algorithm}
\usepackage{algpseudocode}

\newcounter{bla}

\journal{Computer Physics Communications}

\begin{document}

\begin{frontmatter}



\title{An alternative GPU acceleration for a pseudopotential plane-waves density functional theory code with applications to metallic systems}


\author[a,b]{Xuejun Gong\corref{author}}
\author[a,b]{Andrea Dal Corso}

\cortext[author] {Corresponding author.\\\textit{E-mail address:} xgong@sissa.it}
\address[a]{International School for Advanced Studies(SISSA), Via Bonomea 265, Trieste, Italy, 34136}
\address[b]{CNR-IOM, Via Bonomea 265, Trieste, Italy, 34136}

\begin{abstract}
We present an alternative \texttt{GPU} acceleration for plane waves pseudopotentials electronic structure codes designed for systems that have small unit cells but require a large number of {\bf k} points to sample the Brillouin zone as happens, for instance, in metals. We discuss the diagonalization of the Kohn and Sham equations and the solution of the linear system derived in density functional perturbation theory. Both problems take advantage from a rewriting of the routine that applies the Hamiltonian to the Bloch wave-functions to work simultaneously (in parallel on the \texttt{GPU} threads) on the wave-functions with different
wave-vectors {\bf k}, as many as allowed by the \texttt{GPU} memory. Our implementation is written in \texttt{CUDA Fortran} and makes extensive use of kernel routines that run on the \texttt{GPU} (\texttt{GLOBAL} routines) or can be called from inside the \texttt{GPU} threads (\texttt{DEVICE} routines). We compare our method with the \texttt{CPUs} only calculation and with the approach currently implemented in \texttt{Quantum ESPRESSO} that uses \texttt{GPU} accelerated libraries for the \texttt{FFT} and for the linear algebra tasks such as the matrix-matrix multiplications as well as \texttt{OpenACC} directives for loop parallelization. We show in a realistic example that our method can give a significant improvement in the cases for which it has been designed.
\end{abstract}


\end{frontmatter}

\noindent
The published version of this paper can be found at
DOI: \\
https://doi.org/10.1016j.cpc.2024.109439 

\newpage



  \tableofcontents

\section{Introduction}

Density functional theory~\cite{hohenberg_inhomogeneous_1964} (DFT) 
and the availability of more and more powerful computers has made the study of material properties from first-principles a well established reality. Several tools have been refined over the years to solve the one electron Kohn and Sham equations that derive from DFT,~\cite{kohn_self_consistent_1965} the most widespread being based on a plane waves basis and pseudopotentials.
Well tested, freely available~\cite{qe1,qe2,Gonze2020} or commercial~\cite{PhysRevB.59.1758,ClarkSegallPickardHasnipProbertRefsonPayne+2005+567+570} packages implement the theory and allow the calculation of material properties.

In the last ten years, high performance computers aiming to reach the exaflops ($10^{18}$ floating point operations per second) switched to a hybrid technology in which the graphic processing units (\texttt{GPUs}) support the central processing units (\texttt{CPUs}) in the floating point operations. In theory, the \texttt{GPUs} can deliver one or two orders of magnitude more flops than the \texttt{CPUs} themselves, and to harness this power many electronic structure groups are modifying their codes to run on the \texttt{GPUs}.~\cite{gpu_book,QEGPU,smith_portable_2022,cp2k_gpu,abinit_on_gpu,vasp_on_gpu}

Extensions of common programming languages such as \texttt{C/C++} or \texttt{Fortran} 
have provided commands to allocate data on
the \texttt{GPU}, to move data from the \texttt{CPU} to the \texttt{GPU} and vice versa, and to perform  calculations on these data with the \texttt{GPU}. 
\texttt{CUDA Fortran} commands,~\cite{cuda} declarations, and compiler directives and \texttt{OpenACC} compiler directives~\cite{openacc} are two of the most commonly used \texttt{Fortran} extensions. Recently also applications
based on \texttt{openMP} started to appear in the literature.~\cite{datta_accelerating_2023,ruffinoQuantumESPRESSOPerformance2024a} Actually \texttt{OpenACC} and \texttt{OpenMP} have the additional benefit of being transferable to \texttt{GPUs} architectures of different vendors such as \texttt{NVIDIA},  \texttt{AMD}, or \texttt{Intel} and sometimes are preferred to \texttt{CUDA Fortran} that is limited to
\texttt{NVIDIA GPUs}. 

So far, in several plane-waves pseudopotentials and quantum chemistry codes, the \texttt{GPUs} have been exploited by allocating the variables on the \texttt{GPU} and by substituting the calls to linear algebra and fast Fourier transform (\texttt{FFTs}) routines with calls to optimized library routines (such as \texttt{cuBlas},~\cite{cublas} \texttt{cuSolver},~\cite{cusolver} \texttt{cuFFT},~\cite{cuFFT} and \texttt{MAGMA}~\cite{noauthor_magma_nodate}) developed by the \texttt{GPUs} vendors and capable to run on the \texttt{GPU}.~\cite{doi:https://doi.org/10.1002/9781118670712.ch7,QEGPU,datta_accelerating_2023, huhnGPUAccelerationAllelectron2020a, fattebertHybridProgrammingmodelStrategies2024,dasDFTFE10Massively2022}
Sometimes the routines of these libraries have the
same names and arguments of the corresponding \texttt{CPU} libraries and it suffices to allocate the variables on the \texttt{GPU} to call the  \texttt{GPU} routines with minimal changes to the underlying codes and algorithms. Single loops using variables allocated on the \texttt{GPUs} can also be accelerated by compiler directives.
 
In Quantum ESPRESSO~\cite{qe1,qe2} work on this kind of acceleration started more than ten years ago~\cite{spigaPhiGEMMCPUGPULibrary2012a} and has been improved over the years~\cite{romeroPerformanceStudyQuantum2018} leading to a well tested package.~\cite{QEGPU}
Accelerations of \texttt{2X} or higher with respect to the \texttt{CPU} are often found in pseudopotential plane waves codes that adopt this approach.
However, test systems are usually big supercells with many atoms for which the time spent to make calculations on the \texttt{GPU} is larger than the time needed to transfer data from the \texttt{CPU} to the \texttt{GPU}, while small size systems are left out from these tests.  For some applications, metallic systems with small unit cells need tens or hundreds thousands {\bf k} points to sample the Fermi surface.~\cite{mp,gong_pressure_2024,gongInitioQuasiharmonicThermoelasticity2024,malica_quasi-harmonic_2020,malica_quasi-harmonic_2021,thakur_ab_2024,thakur_thermodynamic_2024,dal_corso_elastic_2016,gong_high_2024} The calculation of the phonon dispersions of these systems for many geometries as required for thermodynamic calculations is a problem that could take advantage from the new supercomputers, 
but for small systems we found that the use of the \texttt{GPUs} with the present codes is not always convenient and sometimes it can also slow down the calculation with respect to the \texttt{CPUs} alone.

We have therefore tried to improve the situation and in
this paper we present the solution that we have found: an alternative approach to the acceleration of the pseudopotentials plane waves codes that is useful to deal with metallic system when there are many {\bf k} points. We load on the \texttt{GPU} many wave-functions (i.e. {\bf k} points), all the available ones if the \texttt{GPU} memory is large enough or as many as possible until there is free \texttt{GPU} memory. Then we make the calculations simultaneously on all these data (application of the Hamiltonian to the wave-functions) with each \texttt{GPU} thread working on a single wave-function or on a part of it. To obtain the precise control of the \texttt{GPU} threads that is needed we wrote a set of kernel functions (called \texttt{GLOBAL} in the \texttt{CUDA} language) that implement the theory and run on the \texttt{GPU}. These kernel functions need to call linear algebra and \texttt{FFT} library functions from inside the \texttt{GPU} threads. Unfortunately, libraries such as  \texttt{cuFFT}, \texttt{cuSolver}, or \texttt{MAGMA}, which are called from the \texttt{CPU} and automatically control the number of threads, are not suited for our algorithm. We need functions that can be called 
from \texttt{inside} the \texttt{GPU} threads (\texttt{DEVICE} functions in the \texttt{CUDA} language). We only find the \texttt{C++} library \texttt{cuFFTDx}~\cite{cuFFTDx} that implements such functionalities, but it does not provide a \texttt{FORTRAN} interface so far. For the moment, we transformed to the \texttt{DEVICE} form the \texttt{FORTRAN} sources of \texttt{fftpack5.1}
and of selected \texttt{LAPACK} routines.
Finally, we obtained a code significantly faster than the standard one for small systems with many {\bf k} points.

We start with a brief introduction of the main equations that are solved in a plane-waves pseudopotential code. We stress in particular the algorithms that are relevant for the following discussion, neglecting the parts that have not changed or that are still calculated on the \texttt{CPU}. We then discuss how, in our method, the different parts of the code have been accelerated on the \texttt{GPU}.
Finally, we present a test of our implementation and compare the times required by our approach with those taken by the \texttt{CPUs} only calculations and by the previously available \texttt{GPU} implementation.

\section{Theory}

The solutions of the Kohn and Sham (KS) equations
minimize the DFT total energy. These equations
are an eigenvalue problem for norm conserving pseudopotentials,~\cite{kleinman_efficacious_1982} 
and a generalized eigenvalue problem
for ultrasoft~\cite{Vanderbilt_soft_1990} or projector-augmented wave (PAW) pseudopotentials.~\cite{blochl_projector_1994,kresse_ultrasoft_1999} For periodic solids they can be written as: 
\begin{equation}
{H}_{KS}\psi_{{\bf k}\nu} =\varepsilon_{{\bf k}\nu} S\psi_{{\bf k}\nu},
\label{ks}
\end{equation}
where ${\bf k}$ is a wave vector and $\nu$ is a band index. $H_{KS}$ is the Kohn and Sham Hamiltonian and $S$ is the overlap matrix. We are interested in finding the lowest $N_b$ (number of bands) eigenvalues and eigenvectors of these equations. 
 The Kohn and Sham Hamiltonian depends itself from a potential that is calculated from the charge density (that also depends on the wavefunctions).
It is possible to solve this problem by a self-consistent procedure in which the wavefunctions are first calculated with an approximate potential. Then
these wavefunctions are used to recompute the charge density and a new potential. The latter is mixed with the potential of the previous iterations and the procedure is repeated until one reaches self-consistency.
At each step of the procedure however one has to diagonalize a fixed Hamiltonian which is progressively improved.

In the standard algorithm the problem is solved sequentially for each ${\bf k}$ vector and the charge density is computed at the end when all wave-functions are available.

\subsection{Davidson algorithm}
There are several algorithms currently implemented in electronic structure codes to find the eigenvalues and eigenfunctions in Eq.~\ref{ks}, but here we limit the discussion to the Davidson algorithm.~\cite{davidson_iterative_1975}
In this algorithm an initial set of $N_b$ functions $|\phi_i^{(n)}\rangle$ are progressively improved by enlarging the set applying $H_{KS}-\varepsilon_i S$ and solving the generalized eigenvalue problem $\tilde H_{ij}-\varepsilon \tilde S_{ij}$ on the basis formed by the original and the newly calculated vectors. A standard software library for numerical linear algebra, such as \texttt{LAPACK},~\cite{lapack99} is employed for the diagonalization. The algoritm is the following and has to be repeated for each
{\bf k} point:

\begin{itemize}
\item Given $N_b$ trial eigenpairs: $\quad\left\{\left|\phi_i^{(n)}\right\rangle, \varepsilon_i^{(n)}\right\}$
of the reduced Hamiltonian calculate:
\begin{equation}
\tilde{H}_{i j}=\left\langle\phi_i^{(n)}\left|H_{K S}\right| \phi_j^{(n)}\right\rangle, \quad \tilde{S}_{i j}=\left\langle\phi_i^{(n)}|S| \phi_j^{(n)}\right\rangle.
\end{equation}

\item Build the correction vectors $\left|\tilde{\phi}_i^{(n)}\right\rangle$:
\begin{equation}
\left|\tilde{\phi}_i^{(n)}\right\rangle=\left(H_{d i a g}-\varepsilon_i^{(n)} S_{d i a g}\right)^{-1}\left(H_{K S}-\varepsilon_i^{(n)} S\right)\left|\phi_i^{(n)}\right\rangle,
\end{equation}
where $H_{diag}$ and $S_{diag}$ are the diagonal elements of $H_{K S}$ and $S$ in the plane waves representation.

\item 
Normalize the correction vectors:
\begin{equation}
|\tilde{\phi}_i^{(n)}\rangle={
|\tilde{\phi}_i^{(n)}\rangle \over \sqrt{\langle\tilde{\phi}_i^{(n)}| \tilde\phi_i^{(n)}\rangle}}.
\end{equation}

\item Build an extended reduced Hamiltonian and overlap matrix:
\begin{equation}
\tilde{H}_{i j}=\left\langle\phi_i^{(n)} / \tilde{\phi}_i^{(n)}\left|H_{K S}\right| \phi_j^{(n)} / \tilde{\phi}_j^{(n)}\right\rangle, \quad \tilde{S}_{i j}=\left\langle\phi_i^{(n)} / \tilde{\phi}_i^{(n)}|S| \phi_j^{(n)} / \tilde{\phi}_j^{(n)}\right\rangle.
\end{equation}
\end{itemize}

\begin{itemize}

\item Set $N_{base}$ equal to the number of basis vector. Diagonalize the small $N_{base} \times N_{base}$ reduced Hamiltonian to get the new estimate for the lowest $N_b$ eigenpairs:
\begin{equation}
(\tilde{H}-\varepsilon \tilde{S}) v=0 \quad  \\ \longrightarrow \quad\left\{\left|\phi_i^{(n+1)}\right\rangle, \varepsilon_i^{(n+1)}\right\}.
\end{equation}

\item Calculate $|\tilde{\phi_i}^{(n+1)}\rangle$ for all $i$ for which $\left|\varepsilon_i^{(n+1)}-\varepsilon_i^{(n)}\right|>\varepsilon_{th}$ where $\varepsilon_{th}$ is the accuracy required for the eigenvalues and call $N_{nc}$ the number of new vectors.

\item  If $N_{nc}>0$ repeat with the basis $|\phi_i^{(n)} / \tilde{\phi_i}^{(n)} / \tilde{\phi_i}^{(n+1)} \rangle$ of size $N_{base} + N_{nc}$ and continue with 
progressively larger basis. When the size of the basis becomes too large for the allocated memory instead of adding $\quad\left\{\left|\tilde{\phi_i}^{(n+1)}\right\rangle\right\}$ to the basis,
restart with $\quad\left\{\left|\phi_i^{(n+1)}\right\rangle, \varepsilon_i^{(n+1)}\right\}$.  If $N_{nc}=0$ exit with eigenpairs $\quad\left\{\left|\phi_i^{(n+1)}\right\rangle, \varepsilon_i^{(n+1)}\right\}$
\end{itemize}

The most time consuming step
of this algorithm is the application of the operators
$H_{KS}$ and $S$ to the wave-functions as discussed in the next subsection.

\subsection{Application of Hamiltonian}
The KS Hamiltonian can be written as:~\cite{pickett_pseudopotential_1989}
\begin{equation}
{H_{KS}}\psi_{{\bf k}\nu} = \underbrace{ -\frac{1}{2}\nabla^{2}\psi_{{\bf k}\nu}}_{ kinetic \ energy} 
+\underbrace{V_{eff}\psi_{{\bf k} \nu}}_{ local \ energy}
+\underbrace{V_{NL}\psi_{{\bf k}\nu}}_{non-local \ energy},
\end{equation}
where the effective potentials is the sum of the local, Hartree, and exchange and correlation potentials:
\begin{equation}
V_{eff} = V_{loc}+V_{H}+V_{XC},
\end{equation}
while the nonlocal pseudopotential
is defined in term of the projector functions $\left|\beta^{I}_{m} \right\rangle$
and pseudopotential coefficients
$D^I_{mn}$:~\cite{kleinman_efficacious_1982,Vanderbilt_soft_1990}

\begin{equation}
V_{NL}\left|\psi_{{\bf k}\nu}\right\rangle = \sum_{Imn}{D^I_{mn}}{\left|\beta^{I}_{m} \right\rangle}{\left\langle \beta^{I}_{n} \right|\psi_{{\bf k}\nu} \big \rangle}.
\end{equation}
Here $I$ indicates the different atoms in
the solid and $m$ and $n$ run on all the $\beta^I_m$ functions of a given atom.

The overlap matrix can be calculated in a similar way:~\cite{Vanderbilt_soft_1990}
\begin{equation}
S\left|\psi_{{\bf k}\nu}\right\rangle = \left|\psi_{{\bf k}\nu}\right\rangle+\sum_{mn}q^{I}_{mn}\left|\beta^{I}_{m}\right\rangle\left\langle \beta^{I}_{n} \big| \psi_{{\bf k}\nu} \right\rangle,
\end{equation}
where the coefficients $q^I_{mn}$ are defined together with the pseudopotential.

\subsubsection{Kinetic energy}
The kinetic energy is calculated in reciprocal space. Using the Bloch theorem we write the Bloch wave-functions as:
\begin{equation}
\psi_{{\bf k} \nu}({\bf r})=
e^{i{\bf k}{\bf r}} u_{{\bf k} \nu}({\bf r}) = \frac{1} {\sqrt{V}} \sum_{{\bf G}} C_{{\bf k}+{\bf G}\nu} e^{i ({\bf k}+{\bf G}) {\bf r}},
\end{equation}
where $u_{{\bf k} \nu}({\bf r})$ is a lattice periodic function
expanded in plane waves (here $V$ is the volume of the solid) and the sum is over the reciprocal lattice vectors contained into a sphere defined
by the relationship:
\begin{equation}
{1\over 2}|{\bf k}+{\bf G}|^2 < E_{cut},
\end{equation}
where $E_{cut}$ is the kinetic energy cut-off.
Then we have:
\begin{equation}
-{1\over 2} \nabla^2\psi_{{\bf k} \nu}({\bf r})
= {1 \over \sqrt{V}} \sum_{{\bf G}} C'_{{\bf k}+{\bf G}\nu} e^{i ({\bf k}+{\bf G}) {\bf r}},
\end{equation}
where:
\begin{equation}
C'_{{\bf k}+{\bf G}\nu}=
{1\over 2}|{\bf k}+{\bf G}|^2
C_{{\bf k}+{\bf G}\nu}.
\end{equation}

\subsubsection{Local potential}

The fast Fourier transform (\texttt{FFT}) transforms functions in real space into reciprocal space and the inverse \texttt{FFT} makes the inverse transformation.
\\
From the coefficients $C_{{\bf k}+{\bf G}\nu}$, applying an inverse \texttt{FFT} we obtain the Bloch function in real space (up to a factor $1/\sqrt{V}$):

\begin{equation}
C_{{\bf k}+{\bf G}\nu} \xrightarrow{FFT^{-1}} u_{{\bf k}\nu}({\bf r}) = \sum_{{\bf G}} C_{{\bf k}+{\bf G}\nu} e^{i{\bf G} {\bf r}}.
\end{equation}
The effective potential is applied in real space as:
\begin{equation}
u'_{{\bf k}\nu}({\bf r}) = V_{eff}({\bf r})u_{{\bf k}\nu}({\bf r}),
\end{equation}
and a final \texttt{FFT} computes the 
plane wave expansion of 
$u'_{{\bf k}\nu}({\bf r})$:
\begin{equation}
u'_{{\bf k}\nu}({\bf r})\xrightarrow{{FFT}} C'_{{\bf k}+{\bf G}\nu} = {1 \over N_{\bf r}} \sum_{{\bf r}} u'_{{\bf k}\nu}({\bf r}) e^{-i{\bf G} {\bf r}},
\end{equation}
where $N_{\bf r}$ is the number of points of the \texttt{FFT} grid (see below).

The actual calculation of $V_{eff}$ requires the calculation of the charge density in terms of the wave-functions $\psi_{{\bf k}\nu}$. However since we have not modified this part of the calculation we do not discuss it in detail. We assume only to have a function $V_{eff}$ defined in the points of the \texttt{FFT} grid ${\bf r}$.

\subsubsection{Non local pseudopotential and overlap matrix}
The application of the non local potential needs three matrix-matrix multiplications:

\begin{equation}
\lambda^I_{n {\bf k}\nu} = \left\langle \beta^{I}_{n} \big|\psi_{{\bf k}\nu} \right \rangle = \sum_{\bf G} \beta^{I}_{n}({\bf k}+{\bf G})^{*} C_{{\bf k}+{\bf G}\nu},
\label{lambdah}
\end{equation}

\begin{equation}
\gamma^I_{m {\bf k}\nu} = \sum_{n}{D^I_{mn}}\lambda^I_{n {\bf k}\nu},
\label{gamma}
\end{equation}

\begin{equation}
V_{NL}\left|\psi_{{\bf k}\nu}\right\rangle =\sum_{Im}\gamma^I_{m {\bf k}\nu}\left|\beta^{I}_{m}\right\rangle.
\label{finalnl}
\end{equation}

Similarly, the application of the $S$ matrix is:

\begin{equation}
\delta^I_{m {\bf k}\nu} = \sum_{n}{q^I_{mn}}\lambda^I_{n {\bf k}\nu},
\label{delta_eq}
\end{equation}

\begin{equation}
S\left|\psi_{{\bf k}\nu}\right\rangle =
\left|\psi_{{\bf k}\nu}\right\rangle+
\sum_{Im}\delta^I_{m {\bf k}\nu}\left|\beta^{I}_{m}\right\rangle.
\label{finals}
\end{equation}
where $\lambda^I_{n {\bf k}\nu}$ are those calculated in Eq.~\ref{lambdah}.

\subsection{Density functional perturbation theory}

The phonon frequencies and displacement
modes are obtained by diagonalization of the
dynamical matrix:

\begin{equation}
\omega_{\mathbf{q}}^{2} \mathbf{u}_{s \alpha}(\mathbf{q})=\sum_{s^{\prime} \beta} D_{s \alpha s^{\prime} \beta}(\mathbf{q}) \mathbf{u}_{s^{\prime} \beta}(\mathbf{q}),
\end{equation}
where $D_{s \alpha s^{\prime} \beta}(\mathbf{q})$ is the dynamical matrix:
\begin{equation}
D_{s \alpha s^{\prime} \beta}(\mathbf{q})=\frac{1}{\sqrt{M_{s} M_{s^{\prime}}}} \sum_{\nu} \frac{\partial^{2} E_{t o t}}{\partial \mathbf{u}_{\mu s \alpha} \partial \mathbf{u}_{\nu s^{\prime} \beta}} e^{i \mathbf{q}\left(\mathbf{R}_{\nu}-\mathbf{R}_{\mu}\right)},
\end{equation}
where $E_{tot}$ is the DFT total energy, ${\bf q}$ is a wave vector in the Brillouin zone (BZ),
$\mathbf{R}_{\mu}$ are the Bravais lattice vectors,
$M_s$ are the atomic masses, and $\mathbf{u}_{\mu s \alpha}$ are the atomic displacements.

The second derivative of the DFT total energy can be written in terms of the change of the wave-functions due to a phonon perturbation projected on the
conduction band. These functions are the solutions of a linear system:~\cite{baroni_phonons_2001,dalcorso_density_2001}
\begin{equation}
\left[H_{KS}^{{\bf k}+{\bf q}}+\alpha Q^{{\bf k}+{\bf q}}-\varepsilon_{{\bf k}\nu} S\right] P_c^{{\bf k}+{\bf q}} \frac{\partial u_{{\bf k}\nu}(\mathbf{r})}{\partial {\bf u}_{s'\beta}({\bf q})}=-
P_c^{{\bf k}+{\bf q}} \left[\frac{\partial V_{K S}}{\partial {\bf u}_{s'\beta}({\bf q})} - \varepsilon_{{\bf k},\nu} \frac{\partial S}{\partial {\bf u}_{s'\beta}({\bf q})} \right] u_{{\bf k}\nu}(\mathbf{r}),
\label{linear_system}
\end{equation}
where $P_c^{{\bf k}+{\bf q}}$ is the projector in the conduction band and $\frac{\partial V_{K S}}{\partial {\bf u}_{s'\beta}({\bf q})}=\frac{\partial V_{\text {loc }}}{\partial {\bf u}_{s'\beta}({\bf q})}+\frac{\partial V_H}{\partial {\bf u}_{s'\beta}({\bf q})}+\frac{\partial V_{\text {xc }}}{\partial {\bf u}_{s'\beta}({\bf q})}+
\frac{\partial V_{\text {NL }}}{\partial {\bf u}_{s'\beta}({\bf q})}$. The change of the Hartree
and exchange and correlation potential are:
\begin{equation}
\begin{aligned}
\frac{\partial V_H}{\partial {\bf u}_{s'\beta}({\bf q})} & =\int \frac{e^{i{\bf q}({\bf r}'-{\bf r})}}{\left|\mathbf{r}-\mathbf{r}^{\prime}\right|} \frac{\partial \rho\left(\mathbf{r}^{\prime}\right)}{\partial {\bf u}_{s'\beta}({\bf q})} d^3 r^{\prime}, \\
\frac{\partial V_{x c}}{\partial {\bf u}_{s'\beta}({\bf q})} & =\frac{d V_{x c}}{d \rho} \frac{\partial \rho(\mathbf{r})}{\partial {\bf u}_{s'\beta}({\bf q})},
\end{aligned}
\end{equation}
and depend self-consistently on the charge density induced by the perturbation:
\begin{equation}
\frac{\partial \rho(\mathbf{r})}{\partial {\bf u}_{s'\beta}({\bf q})}= 4 \sum_{{\bf k}\nu} \left[u_{{\bf k}\nu}^*(\mathbf{r}) P_c^{{\bf k}+{\bf q}} \frac{\partial u_{{\bf k}\nu}(\mathbf{r})}{\partial {\bf u}_{s'\beta}({\bf q})}\right]
+  \Delta_{{\bf u}_{s'\beta}({\bf q})} ({\bf r}),
\label{drho}
\end{equation}
where the last term represents the change of the augmentation charge calculated in the
ultrasoft and PAW case but not accelerated
in the present work.~\cite{dalcorso_density_2001} $Q^{{\bf k}+{\bf q}}$ is a projection in the valence manifold,~\cite{baroni_phonons_2001} while the change of the nonlocal pseudopotential is described in more detail in the given references
(see for instance Ref.~\cite{dalcorso_density_2001}).

$\alpha Q^{{\bf k}+{\bf q}}$ can be written in the form:
\begin{equation}
\alpha Q^{{\bf k}+{\bf q}} = \alpha \sum_{\mu}  S |u_{{\bf k}+{\bf q}\mu}
\rangle\langle u_{{\bf k}+{\bf q}\mu} | S,
\label{projector}
\end{equation}
and its application to a set of wave functions $|x_{{\bf k}+{\bf q} \nu j}\rangle$ (here $j$ indicates the different perturbations ${s' \beta}$) can be calculated easily using the fact that $S|x_{{\bf k}+{\bf q} \nu j}\rangle$ is already known from the routine that applies $H^{{\bf k}+{\bf q}}_{KS}$ and $S$. We have a first matrix-matrix multiplication:
\begin{equation}
\mu_{{\bf k}+{\bf q} j \mu\nu} = \langle u_{{\bf k}+{\bf q}\mu} | S |x_{{\bf k}+{\bf q}\nu j}\rangle \alpha,
\label{compute_mu}
\end{equation}
then a second one:
\begin{equation}
|y_{{\bf k}+{\bf q}\nu j} \rangle =\sum_\mu | u_{{\bf k}+{\bf q}\mu}\rangle \mu_{{\bf k}+{\bf q} j \mu\nu},
\label{use_mu}
\end{equation}
and finally we must apply $S$ to the vectors $|y_{{\bf k}+{\bf q}\nu j} \rangle$ and we have:
\begin{equation}
\alpha Q^{{\bf k}+{\bf q}} |x_{{\bf k}+{\bf q} \nu j}\rangle=  S |y_{{\bf k}+{\bf q}\nu j} \rangle=
|y_{{\bf k}+{\bf q}\nu j} \rangle + \sum_{Imn} q^I_{mn}
|\beta^I_m\rangle \langle\beta^I_n|y_{{\bf k}+{\bf q}\nu j} \rangle,
\label{apply_s_ch}
\end{equation}
and this requires other three matrix-matrix multiplications as illustrated above.

The self-consistent linear system (Eq.~\ref{linear_system}) is solved by iterations. From an initial guess of the potentials,
of ${\partial V_H \over {\partial {\bf u}_{s'\beta}({\bf q})}}+{\partial V_{xc} \over {\partial {\bf u}_{s'\beta}({\bf q})}}$, the
linear system is solved and new
induced charge and potentials are obtained. Mixing the latter with the potentials used in the linear system it is possible to reach a self-consistent solution.

The most time consuming step of this process is however the solution of the linear system 
at fixed ${\partial V_H \over {\partial {\bf u}_{s'\beta}({\bf q})}}+{\partial V_{xc} \over {\partial {\bf u}_{s'\beta}({\bf q})}}$ so we will focus on this step.

\subsection{Preconditioned conjugate gradient}

The algorithm used for the solution of Eq.~\ref{linear_system} with a given right hand side is a preconditioned conjugate-gradient iterative algorithm.~\cite{fft_nr,hestenes_methods_1952,fugallo_ab_2013}
Given a starting guess $x$ of the solution of the problem $Ax=b$, we improve it with
the following algorithm:
\begin{eqnarray}
r&\leftarrow&Ax-b, \label{ax1} \\
d &\leftarrow& M^{-1} r, \label{row2} \\
\rho&\leftarrow&d^T r, \label{row3}\\
\gamma&\leftarrow&{\rho \over \rho_{old} },\label{row4} \\
d&\leftarrow&d + \gamma d_{old}, \label{updated}\\
t &\leftarrow& A d, \label{ax2}\\
\lambda &\leftarrow& -{d^T r \over
d^T t}, \label{lambda}\\
x&\leftarrow&x + \lambda d, \label{updatex} \\
r&\leftarrow&r + \lambda t, \label{updater}\\
d_{old}&\leftarrow& d, \label{updatedold}\\
\rho_{old} &\leftarrow& \rho,
\end{eqnarray}
and iterate from Eq.~\ref{row2} until the modulus of $\rho$ is smaller than an input threshold.
Here the arrows indicate that the variables
on the left are substituted with those on
the right.
$r$ is the negative of the residual vector while $d$ contains minus the preconditioned residual in Eqs.~\ref{row2} to \ref{updated} 
and minus the search direction from  Eq.~\ref{updated}.
Eqs.~\ref{row4} and \ref{updated} are executed only from the second iteration onwards. 
The algorithm requires memory sufficient to save the vectors $r$,
$d$, $t$, $d_{old}$ of the same
size of the input vector $x$. $\rho$ and $\rho_{old}$, as well as $\gamma$ and
$\lambda$, are instead scalars.
Moreover, we need two external routines to apply $A$ and $M^{-1}$.
The most time consuming step is the application of the matrix $A$ to $d$. In our case $A=H^{{\bf k}+{\bf q}}_{KS}+\alpha Q^{{\bf k}+{\bf q}}-\varepsilon_{{\bf k}\nu}S$ so again the acceleration rests on the routine that applies $H^{{\bf k}+{\bf q}}_{KS}$ and  $S$ described above. For the preconditioning the following matrix diagonal in reciprocal space $M_{{\bf G},{\bf G}}= MAX(1.0, {|{\bf k}+{\bf q}+{\bf G}|^2 \over 2 \langle \psi_{{\bf k}+{\bf q}\nu}| -{1\over 2} \nabla^2| \psi_{{\bf k}+{\bf q}\nu}\rangle})$ is used and
this vector is passed to the routine. The conjugate gradient algorithm is applied to each ${\bf k}$ point and to each $N_{pe}$ perturbations. The $N_b$
bands of a given ${\bf k}$ point are optimized together but the different {\bf k} points and different perturbations are treated in sequence, one after the other. Only one ${\bf q}$ is calculated in each run.  This algorithm has been used in the last thirty years in Quantum ESPRESSO to solve the linear system. Similar algorithms, with appropriate modifications, can be used also to minimize the total energy and solve the Kohn and Sham equations.~\cite{teter_solution_1989}

\section{\texttt{GPU} optimization}
\begin{figure}[tp]
\caption{Algorithms used in the standard approach and in our optimized \texttt{GPU} approach for the diagonalization of the Hamiltonian.}
\label{fig1}
\vspace*{-\baselineskip}
\begin{minipage}{\columnwidth}
\begin{algorithm}[H]
\caption{\texttt{CPU} and standard \texttt{GPU} diagonalization}\label{alg:old_gpu}
\begin{algorithmic}
\For{\texttt{ik = 1,nks}}    \Comment{\texttt{nks = \#k points per pool}}
        \State \texttt{build $H_{KS}^{\bf k}$ and $S^{\bf k}$ }
        \State \texttt{compute $\varepsilon_{{\bf k}\nu}$ and $\psi_{{\bf k}\nu}$ by Davidson ($N_b$)}
\EndFor
\end{algorithmic}
\end{algorithm}
\end{minipage}
\begin{minipage}{\columnwidth}
\begin{algorithm}[H]
\caption{Optimized \texttt{GPU} diagonalization}\label{alg:new_gpu}
\begin{algorithmic}
\For{\texttt{ikb = 1,nkblock}} \Comment{\texttt{nkblock = \#k points blocks}}
        \State \texttt{build in parallel $H_{KS}^{\bf k}$ and $S^{\bf k}$ on GPU threads ($N_k$)}
        \State \texttt{compute $\varepsilon_{{\bf k}\nu}$ and $\psi_{{\bf k}\nu}$ by Davidson on GPU threads ($N_b\times N_k$)}
\EndFor
\end{algorithmic}
\end{algorithm}
\end{minipage}
\end{figure}

\begin{figure}[tp]
\caption{Flowchart of the routine that applies $H^{\bf k}_{KS}$ and $S^{\bf k}$ to the wavefunctions. Close to
each routine we write the number of threads that are used to run it on the \texttt{GPU}.} 
\label{fig_h}
\begin{center}
\includegraphics[width=8truecm]{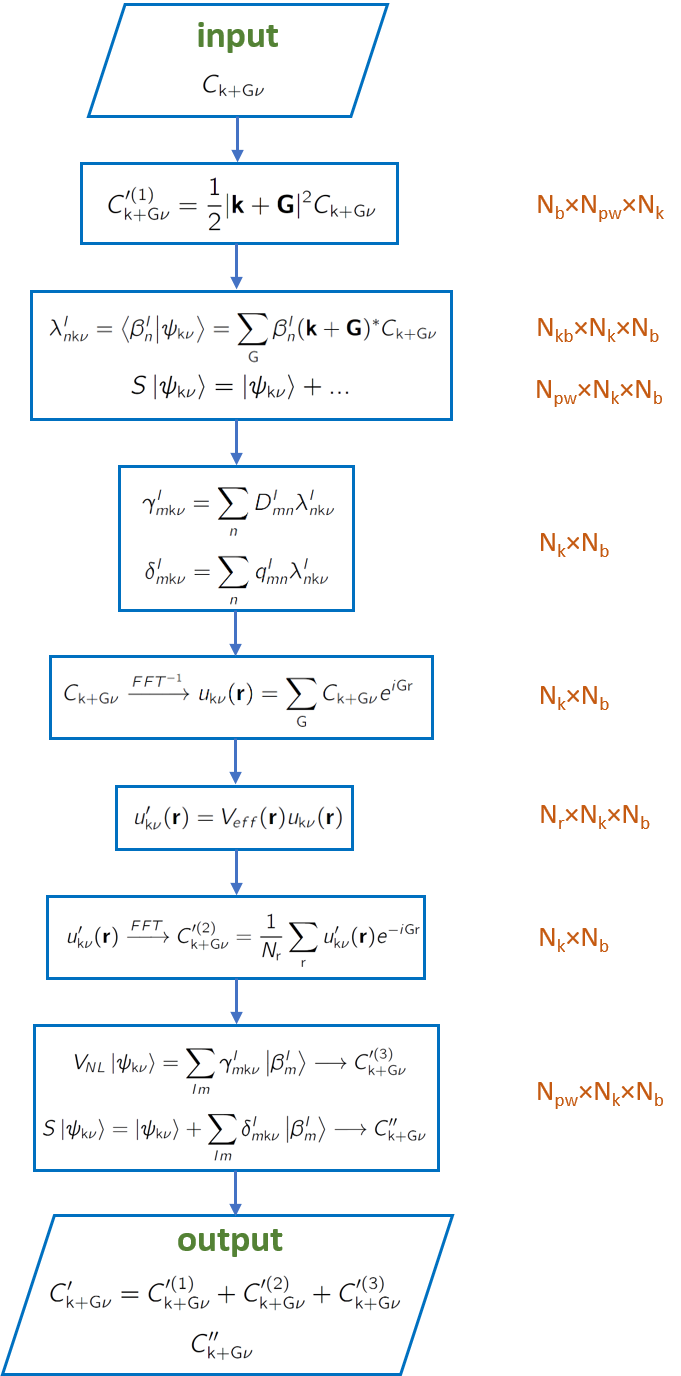}
\end{center}
\end{figure}

\texttt{Quantum ESPRESSO} has several levels of parallelization on the \texttt{CPU}. It is possible to
divide the {\bf k}-points in groups (called \texttt{pools}) and assign each group to a set of 
cores. These cores may be further divided in groups, with each group dealing with a set of bands (\texttt{bands} parallelization) and finally each 
group of cores dealing with a set of bands can 
further divide the
reciprocal lattice vectors ({\bf G}) and work only
on a subset of these (\texttt{{\bf G} vectors} parallelization).
It is at this point that one can introduce the 
\texttt{GPU} acceleration.
The standard method to use the \texttt{GPU} consists into allocating variables on the device memory (the \texttt{GPU}) and to call from \texttt{HOST} routines developed by \texttt{NVidia} that perform linear algebra
(\texttt{cuBLAS}) operations or \texttt{FFTs} (\texttt{cuFFT}) on the data allocated on the device. It is also possible to add compiler directives to run loops in parallel on the \texttt{GPU} without changing the code. 

The standard Davidson algorithm works sequentially on each ${\bf k}$ point of a \texttt{pool} and solves for all the bands (calling the routines that apply $H_{KS}$ and $S$ for a subset of bands, if \texttt{bands} parallelization is used).  Therefore the number of times in which the \texttt{GPU} memory is loaded increases linearly with the number of ${\bf k}$ points. When the size of the problem is small, it can happen that the library matrix-matrix multiplications and \texttt{FFTs} routines cannot exploit all the capacity of the \texttt{GPU} because they have too few data to work on. As a result a \texttt{GPU} calculation might become even slower than a \texttt{CPUs} only calculation.
Parallelizing on the ${\bf G}$ 
vectors just reduces further the size of the 
data allocated on the \texttt{GPU} for each {\bf k} point and does not help in this case. Moreover, presently the \texttt{GPU} acceleration does not work well if many \texttt{CPUs} use the same \texttt{GPU}, so
we use as many \texttt{CPUs} as \texttt{GPUs}.

Our strategy for accelerating the code 
on the \texttt{GPU}
is illustrated schematically in Fig.~\ref{fig1}.
We put on the \texttt{GPU} memory as many wave-functions
(i.e. {\bf k} points) as possible in a block of $N_k$
{\bf k} points and run simultaneously on all these {\bf k} points the operations of the Davidson algorithm needed to diagonalize the Hamiltonian. 
Each \texttt{pool} of CPUs cores works on its
set of {\bf k} points as assigned by the \texttt{pool} parallelization of \texttt{Quantum ESPRESSO} and only these {\bf k} points are divided in blocks for the \texttt{GPU} acceleration. \texttt{Bands} paralellization and {\bf G} vector parallelization presently are not supported by our approach.
The main \texttt{GPU} optimization has been performed on the routine that applies $H_{KS}$ and
$S$ to the wave-functions $\psi_{{\bf k}\nu}$, but some acceleration has been obtained also carrying out the operations of the Davidson algorithm
in parallel on many {\bf k} points. 
In our approach, the routine that applies $H_{KS}$ and $S$
is a \texttt{HOST} routine (i.e. a routine running on the \texttt{CPU}) that receives as input $C_{{\bf k}+{\bf G}\nu}$ for $N_k$ {\bf k} points,
and gives as output the coefficients
$C'_{{\bf k}+{\bf G}\nu}$ and
$C''_{{\bf k}+{\bf G}\nu}$
of the plane waves expansion of $H_{KS}\psi_{{\bf k}\nu}$ and of $S\psi_{{\bf k}\nu}$.
This routine calls in sequence several \texttt{GLOBAL} routines (that is routines that run on the \texttt{GPU} and for which we can specify how many \texttt{GPU} threads run in parallel). The sequence of routines and the formula that they implement is illustrated in Fig.~\ref{fig_h}. The first computes the kinetic energy and runs $N_{pw}\times N_k \times N_b$ threads each one dealing with a ${\bf G}$ vector of one {\bf k} point and of one band (here $N_{pw}$ is the number of ${\bf G}$ vectors used to expand the wave-functions). A second routine computes the scalar product in Eq.~\ref{lambdah} and runs $N_{kb} \times N_k \times N_b$ threads, where $N_{kb}$ is the total number of projectors $|\beta^I_m\rangle$. The latter are loaded on the \texttt{GPU} for all the $N_k$ points before calling the Davidson algorithm. Another \texttt{GLOBAL} routine computes Eqs.~\ref{gamma} and \ref{delta_eq} and runs $N_k \times N_b$ threads, while the sum over $n$ is made inside the routine.
A routine copies $C_{{\bf k}+{\bf G}\nu}$
in $C''_{{\bf k}+{\bf G}\nu}$ and this is made in parallel running $N_{pw}\times N_k \times N_b$ threads. This is the first term of the application of $S$ to the wave functions.
A routine sets to zero the \texttt{FFT} grid running
$N_{\bf r} \times N_k \times N_b$ threads 
and another one sets the non zero elements of this grid running $N_k \times N_b$ threads, each one dealing with all the $N_{\bf r}$ grid points for one ${\bf k}$ point and one band. 
Then a set of three routines applies an inverse \texttt{FFT} to the wave-functions as detailed below, and a routine applies $V_{eff}$ running
$N_{\bf r} \times N_k \times N_b$ threads.
Another set of three routines applies the
\texttt{FFT} to return to reciprocal space and a
routine collects the results from the grid 
and adds them to $C'_{{\bf k}+{\bf G}\nu}$. This is made in parallel on the \texttt{GPU} 
running $N_k \times N_b$ threads. Finally Eq.~\ref{finalnl} is calculated by a routine that runs $N_{pw}\times N_k\times N_b$ threads and adds the result
to $C'_{{\bf k}+{\bf G}\nu}$. In the ultrasoft or PAW PPs case, the same routine calculates also the second term in the right hand side of Eq.~\ref{finals} 
and adds it to $C''_{{\bf k}+{\bf G}\nu}$.

In this algorithm, $N_k$ must be carefully chosen and depends on the amount of \texttt{GPU} memory and on the size of the \texttt{FFT} grid. $N_k$ is mainly limited by the necessity to allocate on the \texttt{GPU} $N_k\times N_b$ \texttt{FFT} grids to apply, in parallel, the local potential to the Bloch functions. The allocation of this memory is done by the \texttt{HOST} routine that implements the Davidson algorithm.

We have also optimized some parts of the 
Davidson algorithm. The standard routine
has been generalized introducing several loops on the $N_k$ {\bf k} points and part of
these loops have been transformed into kernel routines (\texttt{GLOBAL} routines) that perform the calculation in parallel using  $N_k\times N_b$ threads. We have accelerated only the loops that took a significant amount of time. The other loops call the linear algebra \texttt{cuBlas} routines as in the standard approach.

\begin{figure}[tp]
\vspace*{-\baselineskip}
\caption{Algorithms used in the standard phonon code and in our optimized \texttt{GPU} approach for solving the linear system that gives the perturbed wavefunctions.}
\label{fig2}
\begin{minipage}{\columnwidth}
\begin{algorithm}[H]
\caption{\texttt{CPU} and standard \texttt{GPU} phonon algorithm}\label{alg:phonon_old}
\begin{algorithmic}
\For{\texttt{ik = 1,nks}} 
        \State \texttt{build $H_{KS}^{{\bf k}+{\bf q}}$ and $S^{{\bf k}+{\bf q}}$}
    \For{\texttt{ipert = 1,npe}}    \Comment{\texttt{npe = \#perturbations}}  
        \State \texttt{build $P_{c}^{{\bf k}+{\bf q}}\frac{\partial V_{KS}}{\partial u_{s\alpha}({\bf q})} u_{{\bf k}v}$ } ($N_b$)
        \State \texttt{compute $\frac{\partial u_{{\bf k}v}} {\partial u_{s\alpha}({\bf q})}$ by conjugate gradient (CG) ($N_b$)}
    \EndFor
\EndFor
\end{algorithmic}
\end{algorithm}
\end{minipage}
\begin{minipage}{\columnwidth}
\begin{algorithm}[H]
\caption{Optimized \texttt{GPU} phonon  algorithm}\label{alg:phonon_new}
\begin{algorithmic}
\For{\texttt{ikb = 1,nkblock}} \Comment{\texttt{$N_k$ = \#k points per block}}
        \State \texttt{build in parallel $H^{{\bf k}+{\bf q}}_{KS}$ and $S^{{\bf k}+{\bf q}}$ on GPU threads} ($N_k$)
        \State \texttt{build in parallel $P_{c}^{{\bf k}+{\bf q}}\frac{\partial V_{KS}}{\partial u_{s\alpha}({\bf q})} u_{{\bf k}v}$ on GPU threads } ($N_b \times N_k \times N_{pe}$)
        \State \texttt{compute $\frac{\partial u_{{\bf k}v}}{\partial u_{s\alpha}({\bf q})}$ by CG on GPU threads ($N_b \times N_k \times N_{pe}$)}
\EndFor
\end{algorithmic}
\end{algorithm}
\end{minipage}
\end{figure}
The acceleration of the phonon code instead has been carried out essentially on the algorithm that solves the linear system in Eq.~\ref{linear_system}. We
proceed as in the Davidson algorithm (see
the scheme in Fig.~\ref{fig2}). However,
in the phonon case the calculation of the induced charge density (Eq.~\ref{drho}) requires two \texttt{FFT} grids per band, one to contain $u_{{\bf k}\nu}^*(\mathbf{r})$ and one to contain $P_c^{{\bf k}+{\bf q}} \frac{\partial u_{{\bf k}\nu}(\mathbf{r})}{\partial {\bf u}_{s'\beta}({\bf q})}$ so usually we use $N_k$ smaller than in the Davidson algorithm. 
The \texttt{GPU} optimization of the preconditioned conjugate gradient algorithm starts by allocating the \texttt{COMPLEX} vectors $g$, $d$, $d_{old}$, and $t$ on the \texttt{GPU}. For each variable $N_k \times N_b \times N_{pe}$ arrays are allocated. This memory is much larger than the one of the standard algorithm that requires
only $N_b$ copies of each variable, but this space is allocated only on the \texttt{GPU}.
The algorithm is then divided in loops over the $N_k$
{\bf k} points, the $N_{pe}$ perturbations, and the $N_b$ bands.
Loop one executes Eqs.~\ref{row2} and 
Eqs.~\ref{row3}, loop two executes Eq.~\ref{updated}, and loop three computes
$d^T r$ and $d^T t$ that appear in the numerator and denominator of Eq.~\ref{lambda}. Finally loop four computes Eqs.~\ref{updatex}, \ref{updater}, and
\ref{updatedold}. Each loop is transformed 
into routine with the \texttt{GLOBAL} attribute that runs $N_k \times N_b \times N_{pe}$ threads,
each one computing one perturbation to one band of one {\bf k} point. 
Since each thread executes only a scalar
product or an operation of the type $x\leftarrow x + \lambda d$ we have programmed these routines in \texttt{CUDA Fortran} without calling any other library routine. The array themselves instead are not split and each thread works on all the {\bf G} vectors of each wave-function. All the other steps
of the algorithm involve only scalar operations that are performed by the \texttt{CPU}.

Eq.~\ref{ax1} and Eq.~\ref{ax2} require an external routine to apply $A$ to the vectors $x$ (or $d$). For this operator we use the same routine that applies $H_{KS}$ and $S$ in the Davidson algorithm. The routine works in general for an arbitrary number
of wavefunctions so when called from the conjugate gradient algorithm, in parallel on the \texttt{GPU} threads, it deals with the $N_k \times N_{pe}$ set of wavefunctions, each one composed by $N_b$ bands. 
We have then written a \texttt{HOST} routine that 
receives as input the coefficients $C'_{{\bf k}+{\bf q}\nu j}$ and $C''_{{\bf k}+{\bf q}\nu j}$ of the Fourier transform of $H^{{\bf k}+{\bf q}}_{KS}|x_{{\bf k}+{\bf q}\nu j}\rangle$ and $S |x_{{\bf k}+{\bf q}\nu j}\rangle$ and
gives as output the Fourier coefficients of 
$A |x_{{\bf k}+{\bf q}\nu j}\rangle$.
This routine calls a series of \texttt{GLOBAL} routines for which we can control the number \texttt{GPU} threads that run in parallel.
The first routine computes Eq.~\ref{compute_mu} and 
runs $N_k\times N_{pe} \times N_b \times N_b$ threads.
A second routine computes $|a_{{\bf k}+{\bf q}\nu j}\rangle= H_{KS}^{{\bf k}+{\bf q}}|x_{{\bf k}+{\bf q}\nu j}\rangle-\varepsilon_{{\bf k}\nu} S|x_{{\bf k}+{\bf q}\nu j}\rangle$ and runs
$N_k\times N_{pe} \times N_b$ threads.
To complete the operator $A$ we have
to calculate the operator $\alpha Q^{{\bf k}+{\bf q}} |x_{{\bf k}+{\bf q}\nu j}\rangle$ and we optimized also this part to run in many threads on the \texttt{GPU} in parallel on the ${\bf k}$ vectors, the bands, and the perturbations. This is done by calling another set of \texttt{GLOBAL} routines. The first computes Eq.~\ref{use_mu} running on $N_k\times N_{pe} \times N_b$ threads, another one computes
the scalar products $\langle\beta^I_n|y_{{\bf k}+{\bf q}\nu j} \rangle$ that appear in Eq.~\ref{apply_s_ch}
and runs $N_k\times N_{pe} \times N_{kb} \times N_b$ 
threads, and a third routine calculates Eq.~\ref{delta_eq} using the scalar products just calculated and runs
on $N_k\times N_{pe} \times N_b$ threads. Finally
a \texttt{GLOBAL} routine computes Eq.~\ref{apply_s_ch} and adds it to $|a_{{\bf k}+{\bf q}\nu j}\rangle$ running in $N_k\times N_{pe} \times N_b \times N_{pw}$ threads.

\section{Fast Fourier transform}
The application of $V_{eff}({\bf r})$ to one Bloch wave-function requires two Fourier transforms. It is convenient to introduce a mesh in reciprocal space:
\begin{equation}
{\bf G}_{m_1,m_2,m_3}\equiv m_1{\bf b}_1+m_2{\bf b}_2+m_3{\bf b}_3,
\end{equation}
where ${\bf b}_1$, ${\bf b}_2$, and ${\bf b}_3$ are the principal reciprocal lattice vectors and $m_1$, $m_2$, and $m_3$ are integers, 
and a mesh in real space: 
\begin{equation}
{\bf r}_{l_1,l_2,l_3}={l_1 \over N_1} {\bf a}_1 + {l_2 \over N_2} {\bf a}_2 +{l_3 \over N_3} {\bf a}_3, 
\end{equation}
where ${\bf a}_1$, ${\bf a}_2$, and ${\bf a}_3$ are the direct lattice vectors, and $l_1$, $l_2$, and $l_3$ are integers. The integers $N_1$, $N_2$, and $N_3$ define the size of the mesh in real space and, equivalently, the size of the mesh in reciprocal space. They must be sufficiently large so that the vectors ${\bf G}_{m_1,m_2,m_3}$ contain all the vectors ${\bf G}-{\bf G}'$ defined by the basis set.

Given a function in reciprocal space, defined on the {\bf G} vectors $\tilde f(m_1,m_2,m_3) \equiv f({\bf G}_{m_1,m_2,m_3})$, its real space form 
$f(l_1,l_2,l_3)=f({\bf r}_{l_1,l_2,l_3})$ is given by:
\begin{equation}
f(l_1,l_2,l_3)=\sum_{m_1=0}^{N_1-1}\sum_{m_2=0}^{N_2-1}\sum_{m_3=0}^{N_3-1} \tilde f(m_1,m_2,m_3) e^{i2\pi l_1 m_1/N_1} e^{i2\pi l_2 m_2/N_2} e^{i2\pi l_3 m_3/N_3}.
\end{equation}
  
The transform is made in three steps. In the first step, we 
compute $N_1 \times N_2$ one dimensional \texttt{FFTs} along 
$z$:
\begin{equation}
\bar f(m_1,m_2,l_3)=\sum_{m_3=0}^{N_3-1} \tilde f(m_1,m_2,m_3) e^{i2\pi l_3 m_3/N_3}.
\label{zfft}
\end{equation}
We run $N_1 \times N_b \times N_k$ threads on the \texttt{GPU} by calling a \texttt{GLOBAL} routine, and each thread computes $N_2$ \texttt{FFTs}. Each \texttt{FFT} (sum over $m_3$) is carried out by calling an \texttt{FFT} library routine
(\texttt{cfft1b} from \texttt{fftpack.5.1}) which is declared as a \texttt{DEVICE} routine.
In the second step, we compute: 
\begin{equation}
\hat{f} (m_1,l_2,l_3)=\sum_{m_2=0}^{N_2-1}\bar{f}(m_1,m_2,l_3) e^{i2\pi l_2 m_2/N_2}.
\label{yfft}
\end{equation}
In this case we run $N_1 \times N_b \times N_k$ threads each one doing $N_3$ \texttt{FFTs}. Each \texttt{FFT} (sum over $m_2$) is carried out by the \texttt{DEVICE} \texttt{FFT} library routine \texttt{cfft1b}.
Finally, to complete the three dimensional Fourier transform, in the third step we calculate:
\begin{equation}
f (l_1,l_2,l_3)=\sum_{m_1=0}^{N_1-1}\hat{f}(m_1,l_2,l_3) e^{i2\pi l_1 m_1/N_1}.
\label{xfft}
\end{equation}
In this case we run $N_k \times N_b$ threads on the \texttt{GPU} each one computing $N_2 \times N_3$ \texttt{FFTs}. Each one dimensional \texttt{FFT} (sum over $m_1$) is carried out by the \texttt{DEVICE} \texttt{FFT} library routine
\texttt{cfft1b}.
In a similar way one can make a three dimensional Fourier transform to obtain the reciprocal space function from its real space form:
\begin{equation}
\tilde{f} (m_1,m_2,m_3)={1\over N_{\bf r}}\sum_{l_1=0}^{N_1-1}\sum_{l_2=0}^{N_2-1}\sum_{l_3=0}^{N_3-1}
f (l_1,l_2,l_3) e^{-i2\pi l_1 m_1/N_1} e^{-i 2\pi l_2 m_2/N_2} e^{-i2\pi l_3 m_3/N_3},
\label{fwFFT}
\end{equation}
where $N_{\bf r}= N_1 N_2 N_3$.
In this case we call the \texttt{DEVICE} function
\texttt{cfft1f} to actually carry out the one dimensional \texttt{FFTs}.

The product of $V_{eff}$ with the wave-function:
\begin{equation}
u_{{\bf k} \nu}'(l_1,l_2,l_3)=V_{eff}(l_1,l_2,l_3) u_{{\bf k} \nu}(l_1,l_2,l_3),
\end{equation}
is made by running $N_{\bf r} \times N_k \times N_b$ threads on the \texttt{GPU}, each thread computing one product.
After computing the product, an \texttt{FFT} as in Eq.~\ref{fwFFT} gives the Fourier components of the
product that can be added to those obtained by
applying the kinetic energy. This \texttt{FFT} is performed by three routines similar to those described for the inverse \texttt{FFT}.

\subsection{FFT on the device}
Eqs.~\ref{zfft},\ref{yfft},\ref{xfft} cannot be implemented as written since they involve
$N_i^2$ operations, where $N_i$ is $N_1$, $N_2$ or $N_3$. These sums can be done more efficiently with an \texttt{FFT} algorithm that requires $N_i log (N_i)$ operations.~\cite{fft_nr} 
The \texttt{FFTXlib} of \texttt{Quantum ESPRESSO}
contains both the three dimensional \texttt{FFT} driver and a copy of an old \texttt{FFTW} library.~\cite{FFTW05} It also supports the newer \texttt{FFTW3} library, some vendor-specific \texttt{FFT} libraries, and it can call library routines optimized for the \texttt{GPU} in \texttt{cuFFT}~\cite{QEGPU,cuFFT}. Moreover, it can carry out the \texttt{FFT} in parallel when the \texttt{FFT} mesh (and {\bf G} vectors) are distributed among different \texttt{MPI} processes.
 However, these routines are called from the \texttt{CPU HOST} with actual argument variables that are allocated on the \texttt{GPU} and they take care of launching the kernel threads on the \texttt{GPU}.
In our approach, the \texttt{FFT} routines are called from inside the \texttt{GPU} threads and therefore must have the \texttt{DEVICE} attribute, hence \texttt{FFTXlib} cannot be used. The  library that offers this functionality  \texttt{cuFFTDx} is written in \texttt{C++} and it has not yet a \texttt{FORTRAN} interface.
Therefore, we have taken the \texttt{fftpack5.1}~\cite{fftpack} which is distributed under the GNU GPL licence together with its \texttt{Fortran} source and we have modified each routine and function of this library by adding the \texttt{ATTRIBUTES(DEVICE)}.
We have also constructed a \texttt{Fortran} interface for each routine so that the routines that include the interface can know that the routines of \texttt{fftpack} are actually \texttt{DEVICE} routines and accept variables allocated on the \texttt{GPU}.
The modified library is distributed together with
the \texttt{thermo\_pw} package.

\section{Matrix Diagonalization}
The Davidson algorithm requires the solution
of a generalized eigenvalue problem in a reduced basis: 
\begin{equation}
Ax = \lambda Bx,
\end{equation}
where $A$ and $B$ are Hermitian matrices and $\lambda$ and $x$ are the eigenvalues and eigenvectors.
Usually, the \texttt{CPU} makes this calculation by calling \texttt{LAPACK} routines~\cite{lapack99} such as \texttt{ZHEGVX} that computes selected eigenvalues and, optionally, eigenvectors of a complex generalized Hermitian-definite eigenproblem,
or \texttt{ZHEGV} that computes all eigenvalues and eigenvectors of the same matrices. It is also possible to call library routines optimized for the \texttt{GPU} of the \texttt{cuSolver} or of the \texttt{MAGMA} libraries. A \texttt{HOST} driver that calls these routines is contained in the \texttt{LAXlib} library distributed with \texttt{Quantum ESPRESSO}. We have tested this approach creating a loop over the $N_k$ {\bf k} points
that calls these routines, but found that it is possible to obtain a significant speed up by simultaneously diagonalizing the generalized eigenvalue problem for many {\bf k} points. We run therefore a \texttt{GLOBAL} routine with as many threads as possible (ideally $N_k$, but see below). 
 In order to solve the generalized eigenvalue problem inside a \texttt{GLOBAL} routine we cannot call \texttt{HOST} routines such as those available in \texttt{cuSolver} or in \texttt{MAGMA}~\cite{noauthor_magma_nodate} what is needed is a library that can be called from inside the \texttt{GPU} threads
(with \texttt{DEVICE} routines). Since
we are not aware of any \texttt{DEVICE} implementation of \texttt{LAPACK}, we took the routines \texttt{ZHEGVX} and \texttt{ZHEGV} together with those called by them, transformed them into \texttt{DEVICE} routines, and wrote the corresponding \texttt{Fortran} interfaces. 
We found only one problem with this approach: 
The routine \texttt{ZPOTRF2}, which performs the Cholesky factorization of a Hermitian positive definite matrix $A$, is recursive. 
Since \texttt{CUDA Fortran} does not allow for
recursive \texttt{DEVICE} routines or functions, we rewrote it with a non recursive algorithm.  

The number of {\bf k} points that can be diagonalized simultaneously is usually lower than $N_k$ since the \texttt{LAPACK} \texttt{DEVICE} routines use a certain amount of \texttt{GPU} resources. 
So we divided the $N_k$ {\bf k} points in blocks of maximum size determined empirically on the available
machine.

\section{Results}

\subsection{Benchmark Example}

We have implemented our approach in the \texttt{thermo\_pw} code~\cite{dal_corso_thermo_pw_2022}
which is a driver of \texttt{Quantum ESPRESSO} routines to calculate materials properties. To activate the new approach, it suffices to set the flag
\texttt{many\_k} to \texttt{.TRUE.}
and the input variable \texttt{memgpu} to the amount of \texttt{GPU} memory (in GBytes).
Both variables are written in the \texttt{thermo\_control} input file.
The new routines are in the directory \texttt{qe} of \texttt{thermo\_pw}, while the \texttt{LAPACK} and \texttt{fftpack5.1} routines modified with the \texttt{ATTRIBUTES(DEVICE)}
together with their interfaces are distributed in separate subdirectories of the \texttt{thermo\_pw}
package. For further details please refer to the \texttt{thermo\_pw} user's guide.

Our benchmark is a part of the calculations carried out to compute the quasi-harmonic temperature dependent elastic constants of tungsten.~\cite{gong_pressure_2024} 
Our system is body centered cubic (bcc) tungsten
simulated with the PBEsol exchange and correlation functional~\cite{pbesol} at the lattice constant $a=5.965$ a.u.. Tungsten is described with a PAW pseudopotential that has $14$ valence electrons and we compute $N_b=11$ bands.~\cite{corso_dalcorsopslibrary_2022}
We use cut-offs for the wave-functions/charge density of $90$/$360$ Ry, a {\bf k}-point mesh of $45\times45\times45$ and deal with the Fermi surface with the smearing approach (\cite{mp}) with a smearing parameter $\sigma = 0.02$ Ry. 
 The \texttt{FFT} mesh has size $32\times 32 \times 32$ for a total of $N_{\bf r}=32768$ mesh
points.
We compute the phonon frequencies for the point
${\bf q}={2\pi \over a} (-1/8, -1/4, 3/8)$.
The small space group of this ${\bf q}$ point has
no rotational symmetry in it, so we need to use the complete mesh of $45^3=91125$ ${\bf k}$ points when
computing the perturbed wave-functions. Since we need also the eigenvalues and eigenfunctions at ${\bf k}+{\bf q}$ we compute the band structure of $182250$ ${\bf k}$ points.

We report the time obtained with \texttt{version 7.3} of {\texttt{Quantum ESPRESSO} together with {\texttt{thermo\_pw version 2.0.0}}.
All tests have been performed on the \texttt{Leonardo} supercomputer at \texttt{CINECA}. 
Each node of the machine has a \texttt{CPU} with $32$ cores and $4$ Amp\`ere \texttt{GPUs}.
 In the \texttt{Leonardo} manual, the theoretically declared peak performance of one node ($32$ cores) is $1680$ Gflops while the four \texttt{GPUs} of one node can provide $75000$ Gflops. There is therefore a maximum theoretical acceleration of a factor of $45$.
We run on the \texttt{GPUs} using as many \texttt{CPU} cores as \texttt{GPUs} and each \texttt{CPU} runs one \texttt{MPI} process.~\cite{mpi40} \texttt{MPI} processes can communicate among themselves with \texttt{MPI} library calls. Each \texttt{MPI} process communicates with one \texttt{GPU}, multiple \texttt{MPI} processes using the same \texttt{GPU}
are not allowed. Moreover we do not use direct \texttt{GPU}-\texttt{GPU} communication.
When several \texttt{MPI} processes run, the total number of {\bf k} points is divided in a number of
pools equal to the number of \texttt{MPI} processes.
The code is compiled with the \texttt{PGI} \texttt{Fortran} compiler contained in the \texttt{Nvidia SDK}~\cite{gpu_compiler}.

\subsection{FFT}

\input{table1.tex}
In Table~\ref{tab:1}, we report the time necessary to compute the \texttt{FFTs} to apply the local potential.
We consider three cases: \texttt{CPUs} only, standard  \texttt{GPU} code that calls the \texttt{cuFFT} library, and the optimized \texttt{GPU} code that uses the \texttt{fftpack.5.1} routines declared as \texttt{DEVICE} routines. 
In the \texttt{GPUs} runs, we consider $1$, $2$, $4$ or $8$ \texttt{GPUs}. For the \texttt{CPUs} only runs, we use all the \texttt{CPUs} of one ($32$) or two nodes ($64$). Further, with $32$ cores, the
{\bf k} points are divided into $8$, $16$, or $32$
pools, with $64$ cores, into $32$ or $64$ pools. 
When comparing \texttt{CPUs} and \texttt{GPUs}, we compare $4$ or $8$ \texttt{GPUs} with the best times obtained with $32$ or $64$ cores, respectively.
We start by discussing the \texttt{CPUs} only case. With
both $32$ or $64$  cores, the minimum \texttt{FFT} time is obtained when the number of pools
is equal to the number of cores. This indicates
that in this system it is not useful to divide the ${\bf G}$ vectors among
\texttt{CPUs}. The second observation is that when we pass from one to two nodes the time halves, showing a good scaling with the number of nodes. 
We call $T_{cpu}$ the best time obtained with one or
two nodes. 
Passing now to the \texttt{GPU} times, we see that both with the standard algorithm and with the optimized one the computational time is inversely proportional to  the number of \texttt{GPUs}.
Comparing now the time taken by the standard \texttt{GPU} algorithm, we see that
it is $0.76$ T$_{cpu}$ ($4$ \texttt{GPUs}), $0.79$ T$_{cpu}$ ($8$ \texttt{GPUs}). So, as far as the \texttt{FFT} is concerned, it is convenient to use the \texttt{GPUs} instead of the \texttt{CPUs} although the gain is not big. The optimized \texttt{GPU} algorithm gives times that are $0.46$ T$_{cpu}$ ($4$ \texttt{GPU}), $0.42$ T$_{cpu}$ ($8$ \texttt{GPUs})}. This is much less than the theoretical capacity of the \texttt{GPU}, but still it makes convenient to use the latter.
 When computing the \texttt{FFT}, the optimized \texttt{GPU} algorithm is $1.9$ times faster ($8$ \texttt{GPUs}) than the standard \texttt{GPU} version
that calls the \texttt{CuFFT} routines in sequence on the {\bf k} points.

\subsection{Diagonalization}

\input{table2.tex}
In Table~\ref{tab:2}, we report the time spent by the diagonalization of the reduced Hamiltonian carried out by the \texttt{LAPACK} routines on the \texttt{CPU}, by the \texttt{cuSolver} library running on \texttt{GPU} called by the \texttt{LAXlib} package, and by the optimized \texttt{GPU} version of the code in which the Hamiltonians of many {\bf k}-vectors are diagonalized simultaneously by the \texttt{LAPACK} routines declared as \texttt{DEVICE} routines. The effect of using pools and several \texttt{CPUs} is also illustrated.
 The sizes of the matrices to be diagonalized vary, depending on the istantaneous size of the basis set in the Davidson algorithm. The routine must find the lowests $N_b=11$ eigenpairs in a matrix that can have a maximum size equal to $4 N_b=44$. This is repeated for all {\bf k} points for all Davidson iterations and all self-consistent iterations in addition to a band structure calculation before the phonon calculation (in which there are 
about $2 \times 10^5$ {\bf k} points).
The \texttt{CPU} diagonalization time scales
linearly with the number of ${\bf k}$ points and
therefore depends only on the number of pools.
Using $64$ or $32$ cores gives exactly the same time when we use $32$ pools, but if we use a number of
pools equal to the number of cores with two nodes we
halves the diagonalization time with respect to one
node. A good scaling is also shown by the \texttt{GPU}
calculation. Increasing the number of \texttt{GPUs} increases the number of pools and therefore decreases the number of ${\bf k}$ points per \texttt{pool}.
With both the standard \texttt{GPU} algorithm and with the optimized one we could not run faster
than the \texttt{CPU}. With the standard 
algorithm the size of the matrix to diagonalize is so small that the time to initialize the \texttt{GPU} greatly exceeds the \texttt{CPU} diagonalization time.
In this particular example, the time of the standard \texttt{GPU} calculation is $18$ T$_{cpu}$. With our optimization we could reduce this time to $3$ T$_{cpu}$. In our example however the total time for the diagonalization is small with respect to all other times and we have not tried to further optimize this part. 

\subsection{Application of the Hamiltonian and of S}

\input{table3.tex}
In Table~\ref{tab:3} we show the time required
for the application of the Hamiltonian and of the overlap matrix $S$ to the wave-functions. This time comprises the time needed to apply the \texttt{FFT} and inverse \texttt{FFT} to the wave-functions, the time needed to apply the kinetic energy and the nonlocal pseudopotential as
well as the time needed to apply the overlap matrix $S$. In the same table we report also the difference between these times and the times needed to carry out the \texttt{FFT} reported in Table~\ref{tab:1}. 
 In the time reported in the table, we apply the operator $H_{KS}$ and $S$ about $8 \times 10^7$ times (as reported by the code when we do not use the optimized algorithm and $7.3 \times 10^4$ when we use the optimized algorithm and many {\bf k}-points are calculated concurrently). This is reasonable since we have $1 \times 10^5$ {\bf k} points, about $16$ self-consistent iterations and $N_{pe}=3$ modes. This gives an average of $16$ conjugate gradient steps per iteration.
To count the number of operations is more difficult since the number of bands is not always constant.
If take as an average value $N_b=11$ bands,
the number of plane waves $N_{pw}=2093$ and a number of projector functions $N_{kb}=18$ we see that
Eq.~\ref{lambda} is the multiplication of a matrix
$18\times 2023$ and a matrix $2023 \times 11$.
We start by considering the \texttt{CPU} times when \texttt{FFT} time is subtracted. These times depend on the number of cores, but less on how these cores are distributed between ${\bf G}$ vectors and ${\bf k}$-point pools.
Still using only ${\bf k}$-point pools gives the shortest times but the differences are
small. Comparing with the standard \texttt{GPU} version, we see that the application of the nonlocal potential and of the $S$ matrix require too many small size matrix-matrix multiplications and this part of the calculation is quite slow on the \texttt{GPU}. For this calculation the required time is $6$ T$_{cpu}$.
The optimized \texttt{GPU} algorithm is much faster and needs about  $0.42$ T$_{cpu}$. Adding also the speedup obtained with the \texttt{FFT}, the optimized \texttt{GPU} algorithm takes about $0.44$ T$_{cpu}$. Comparing the two \texttt{GPU} algorithms,  the optimized one is $4$ times faster in applying $H_{KS}$ and $14$ times faster in applying the nonlocal pseudopotential and the $S$ matrix.

\subsection{Total time}
\input{table5.tex}
\input{table6.tex}
\input{table7.tex}

In this section we present some benchmarks of the entire run, considering both the self consistent
and the phonon frequencies calculations.
We report in Tables~\ref{tab:5},~\ref{tab:6},~\ref{tab:7} the total time. This time is approximately twice the time required by the application of $H_{KS}$ and $S$ in the \texttt{CPU} and in the optimized \texttt{GPU} cases and three times in the standard \texttt{GPU} case. 
Considering now the total T$_{cpu}$, we see that the 
faster time is obtained when the number of pools is
equal to the number of cores. The scaling with the number of nodes is good: from $32$ to $64$ cores the code is $1.8$ times faster.
The standard \texttt{GPU} approach takes $3.4$ T$_{cpu}$ (one node) or $3.1$ T$_{cpu}$ (two nodes), while the optimized \texttt{GPU} approach takes  $0.58$ T$_{cpu}$ (one node) and $0.48$ T$_{cpu}$ (two nodes). The difference between one and two nodes is due to the different number of \texttt{CPUs} cores available in the two cases. The parts that are not accelerated are calculated faster when more cores are available. Comparing now the two \texttt{GPU} algorithms we see
that the optimized one is about $6$ times faster.

In the table we have indicated also the cost of each run in core-hours. This cost is obtained by multiplying the total time by the number of core
used (in the \texttt{GPU} case, each \texttt{GPU}
costs $8$ cores). We have also added the time needed with $4$ and $8$ nodes ($128$ and $256$ cores).
We find that increasing the number of nodes the total cost tend to increase (even if there are some fluctuations) since it is difficult to achieve an exact linear scaling with the number of \texttt{pools}. In the optimized \texttt{GPU} case the optimum is obtained with $2$ nodes.
It is therefore convenient to carry out this calculations with a small number of nodes per ${\bf q}$ point and calculate in parallel on different nodes different ${\bf q}$ points and geometries. However, even with an ideal scaling with the \texttt{pools} and a computer that can
provide as many \texttt{GPUs} as desired, it is still 
convenient to use pools that contain a number
of ${\bf k}$ point sufficient to occupy the \texttt{GPU} memory and use it (gaining about a factor \texttt{2X}), than split the calculations so that each \texttt{pool} has a single {\bf k} point.

\section{Conclusions and perspectives}

We discussed a scheme to accelerate on the \texttt{GPUs} electronic structure codes based on plane waves and pseudopotentials. We have shown in the example of bcc tungsten that our scheme can be faster than the currently implemented \texttt{GPU} version when the system has small unit cells but requires a thick mesh of {\bf k} points. The main idea is to apply the Hamiltonian to the wave-functions in parallel on many ${\bf k}$ points, one per \texttt{GPU} thread, so as to increase both the size of the data on which the \texttt{GPU} works at any given time and to give to the \texttt{GPU} a sufficient numerical workload to exploit all its SMs.
Our method has been implemented in \texttt{CUDA Fortran} by partially rewriting the code and by using \texttt{GLOBAL} and \texttt{DEVICE} routines to parallelize the work of different \texttt{GPU} threads. We have discussed in detail the optimization of the Davidson algorithm, the application of the Kohn and Sham Hamiltonian and of the overlap matrix $S$ to the wave-functions, and the preconditioned conjugate gradient algorithm which is used to solve the linear system of DFPT. In our example the application of $H_{KS}$ and $S$ to the wave-functions accounts for about one half of the total time with \texttt{CPUs} and
about $1/3$ with the \texttt{GPUs}. For these operations our optimized \texttt{GPU} method is about $6$ times faster than the standard \texttt{GPU} approach, and about twice as fast than the \texttt{CPUs} only calculation. The main limitation of the present implementation is that it does not support the reciprocal lattice vectors distribution among \texttt{CPUs}. It is instead possible to divide the {\bf k} points in pools so that different \texttt{GPUs} acts on different pools. 
Finally, we underline the fact that when the system (and the \texttt{FFT} mesh) becomes large enough the \texttt{cuFFT} library routines become more efficient than our \texttt{DEVICE} routines and at that point the standard approach might become more convenient.

Our approach required a precise control of the \texttt{GPU} threads and math libraries (with \texttt{DEVICE} functions) that can be called from inside the \texttt{GPU} threads. Presently not many libraries offer this functionality and we hope that, in future, optimized \texttt{DEVICE} versions of math libraries will appear together with \texttt{FORTRAN} compatible interfaces. The substitution of our transformed routines with better optimized ones could further improve the speed of our code.
As a last consideration we might ask if there are other ways to speed up the plane-waves pseudopotential codes for metallic cases as those that we need for our research. There are several option that one might explore from introducing a batched form of the \texttt{FFT} and of the linear algebra routines, to using new \texttt{FFT} \texttt{GPU} libraries such as \texttt{heFFTe}.~\cite{heFFTe}. If these options would solve the problem pointed out in this paper within the standard \texttt{GPU} scheme remains still to be investigated.

The implemented software is distributed within the
GPL licence within the \texttt{thermo\_pw} package.~\cite{dal_corso_thermo_pw_2022}  

\section{Appendix: \texttt{GPU} and 
\texttt{CUDA Fortran}}

The \texttt{CUDA} architecture is built around a scalable array of multithreaded Streaming Multiprocessors (SMs). Each SM has a set of execution units and a set of registers and can operate on variables contained in the \texttt{GPU} memory. \texttt{CUDA Fortran} allows the allocation of data on the \texttt{GPU} (called \texttt{DEVICE} in this context), the transfer of data from and to the \texttt{GPU} and the writing of routines that can run on the \texttt{GPU} in many different threads, each one working on different data (called \texttt{GLOBAL}) or that
can be called from the \texttt{GPU} threads (called \texttt{DEVICE}).  This is possible also with \texttt{OpenACC} and \texttt{openMP} compiler directives but
we have opted for \texttt{CUDA Fortran} since 
in this moment there is a large basis of installed supercomputers equipped with \texttt{NVIDIA GPUs} that
can run code written in \texttt{CUDA Fortran} and also the one available to us is in this category. 
The use of \texttt{OpenACC} and \texttt{openMP} that
could be required to make our code transferable to \texttt{GPUs} of other vendors might be considered
in the future if necessary.

In \texttt{CUDA Fortran}, to run on the \texttt{GPU}, one declares the routines with \texttt{ATTRIBUTES(GLOBAL)} or \texttt{ATTRIBUTES(DEVICE)}. Both run on the \texttt{GPU}, but the first can be called from the \texttt{CPU} host with the triple chevron syntax $(<<<,>>>)$ to specify 
the number of threads blocks and threads per block that are employed. In general thread blocks can be arranged in a three dimensional grid with variable size in each dimension. The second can be called from the \texttt{GLOBAL} routines on the data already selected for the current thread.
\texttt{CUDA} makes four pieces of information available to each thread:
the thread index (\texttt{threadIdx}),
the block index (\texttt{blockIdx}),
the size and shape of a block (\texttt{blockDim}), and the size and shape of a grid (\texttt{gridDim}).
This information can be used to choose the variables the current thread will work on. 

To give an order of magnitude, the Volta (Amp\`ere) \texttt{GPU} architecture has $84$ ($108$) SMs each capable of running up to $32$ threads that is $32\times 84=2688$ ($3456$) threads can run simultaneously on the \texttt{GPU}, however each thread block must run the same instructions on different data, while different thread blocks can execute different instructions. The code is independent from the \texttt{GPU} architecture on which it will run and can require even bigger grids and block sizes whose threads are run in sequence on the available SMs.

\section{Acknowledgments}

This work has been supported by the Italian MUR (Ministry of University and Research) through the National Centre for HPC, Big Data, and Quantum Computing (grant No. CN00000013). Computational facilities have been provided by SISSA through its Linux Cluster, ITCS, and the SISSA-CINECA 2021-2024 Agreement. 
Partial support has been received by MAX ``MAterials design at the eXascale" Centre of Excellence 
for Supercomputing applications (Grant agreement No. 
101093374, co-funded by the European High Performance 
Computing joint Undertaking (JU) of the European Union and participating countries). 
X. Gong acknowledges the support received in the framework of the Joint Research Agreement for Magnetic Confinement Fusion between \texttt{Eni} and \texttt{CNR}.

\bibliographystyle{elsarticle-num}
\bibliography{CPC}

\end{document}

%% file: table1.tex
\begin{table}[ht]
\centering
\caption{Comparison of the time spent by computing the FFT and the inverse \texttt{FFT} when applying the Hamiltonian operator in the Davidson algorithm and in the conjugate gradient algorithm for the example described in the paper.}
\vspace{8pt}
\begin{threeparttable}
\scriptsize{
\setlength{\tabcolsep}{2pt}
\begin{tabular}{cccccccccccccc}
\toprule
\toprule
  & \multicolumn{5}{c}{CPU}  & \multicolumn{4}{c}{GPU}  & \multicolumn{4}{c}{optimized GPU} \\

\cmidrule(lr){2-6}  \cmidrule(lr){7-10}  \cmidrule(lr) {11-14}
\#CPU & 32 & 32 & 32 &64 &64 & 1 & 2 & 4 & 8 & 1 & 2 & 4 &8  \\ 
\#GPU & 0 & 0 & 0 & 0& 0&  1 & 2 & 4 &8 & 1 & 2 & 4 &8 \\ 
\#task(np) & 32 & 32 & 32 & 64&64 & 1 & 2 & 4 &8 & 1 & 2 & 4 &8 \\
\#pool(nk)  & 8 & 16 & 32 &32 & 64 &1 & 2 & 4 &8 & 1 & 2 & 4 & 8\\
\midrule
time (s) & 6771 & 6280 &4539  &3269 &2213 & 13415 & 6866& 3441& 1744&  7532 & 3574 &  2068 & 934 \\ 
\bottomrule
\end{tabular}
}
\end{threeparttable}
\label{tab:1}
\end{table}

%% file: table2.tex
\begin{table}[ht]
\centering
\caption{Comparison of the time spent to diagonalize the reduced Hamiltonian using linear algebra routines
(within the Davidson algorithm).}
\vspace{8pt}
\begin{threeparttable}
\scriptsize{
\setlength{\tabcolsep}{2pt}

\begin{tabular}{cccccccccccccc}
\toprule
\toprule
  & \multicolumn{5}{c}{CPU}  & \multicolumn{4}{c}{GPU}  & \multicolumn{4}{c}{optimized GPU} \\

\cmidrule(lr){2-6}  \cmidrule(lr){7-10}  \cmidrule(lr) {11-14}
\#CPU & 32 & 32 & 32 &64 &64 & 1 & 2 & 4 & 8 & 1 & 2 & 4 &8  \\ 
\#GPU & 0 & 0 & 0 & 0& 0&  1 & 2 & 4 &8 & 1 & 2 & 4 &8 \\ 
\#task(np) & 32 & 32 & 32 & 64&64 & 1 & 2 & 4 &8 & 1 & 2 & 4 &8 \\
\#pool(nk)  & 8 & 16 & 32 &32 & 64 &1 & 2 & 4 &8 & 1 & 2 & 4 & 8\\
\midrule
time (s) & 81 & 44 & 20 &21 &10 & 1286 &684 & 348 & 178& 220 & 118 & 60 &29 \\ 
\bottomrule
\end{tabular}
}
\end{threeparttable}
\label{tab:2}
\end{table}

%% file: table3.tex
\begin{table}[ht]
\centering
\caption{Comparison of the total time spent to apply  $H_{KS}$ and $S$ to the wave-functions
in the Davidson algorithm and in the conjugate gradient algorithm.}
\vspace{8pt}
\begin{threeparttable}
\scriptsize{
\setlength{\tabcolsep}{2pt}
\begin{tabular}{cccccccccccccc}
\toprule
\toprule
  & \multicolumn{5}{c}{CPU}  & \multicolumn{4}{c}{GPU}  & \multicolumn{4}{c}{optimized GPU} \\

\cmidrule(lr){2-6}  \cmidrule(lr){7-10}  \cmidrule(lr) {11-14}
\#CPU & 32 & 32 & 32 &64 &64 & 1 & 2 & 4 & 8 & 1 & 2 & 4 &8  \\ 
\#GPU & 0 & 0 & 0 & 0& 0&  1 & 2 & 4 &8 & 1 & 2 & 4 &8 \\ 
\#task(np) & 32 & 32 & 32 & 64&64 & 1 & 2 & 4 &8 & 1 & 2 & 4 &8 \\
\#pool(nk)  & 8 & 16 & 32 &32 & 64 &1 & 2 & 4 &8 & 1 & 2 & 4 & 8\\
\midrule
time (s) & 7907 & 7319 & 5519 & 3780 & 2702 &36157 & 18465 & 9263 & 4694 &  9138 & 4375 & 2472 & 1138 \\ 
time -time$_{\rm FFT}$ (s) & 1136 & 1039 & 980 & 511 & 489 & 22742& 11599 & 5822 & 2950 & 1606 & 801 & 404 & 204 \\
\bottomrule
\end{tabular}
}
\end{threeparttable}
\label{tab:3}
\end{table}

%% file: table5.tex
\begin{table}[ht]
\centering
\caption{Total time spent in the standard \texttt{CPU} calculations. The number of \texttt{CPUs}, \texttt{GPUs}, \texttt{tasks}, and \texttt{pools} are also indicated.
The number of core-hours is obtained multiplying the total time by the number of cores.}
\vspace{8pt}
\begin{threeparttable}
\scriptsize{
\setlength{\tabcolsep}{2pt}
\begin{tabular}{cccccccc}
\toprule
\toprule
  & \multicolumn{7}{c}{CPU} \\
\midrule
\#CPU & 32 & 32 & 32 &64 &64 & 128 & 256\\ 
\#GPU & 0 & 0 & 0 & 0& 0 & 0 & 0 \\ 
\#task(np) & 32 & 32 & 32 & 64&64 & 128 & 256 \\
\#pool(nk)  & 8 & 16 & 32 &32 & 64 & 128 & 256 \\
\midrule
time (s) & 11400 & 10560 & 8040 & 5640& 4500 & 2280  &  1182 \\
time (m) & 190 & 176 & 134 & 94& 75 & 38 & 20 \\
core-hours &  101 & 94 &  71 &  100&  80 & 81 & 85\\
\bottomrule
\end{tabular}
}
\end{threeparttable}
\label{tab:5}
\end{table}

%% file: table6.tex
\begin{table}[ht]
\centering
\caption{Total time spent in the standard \texttt{GPU} calculations. This includes also the time passed on the part of the code that are not GPU accelerated  or are not GPU optimized. The number of \texttt{CPUs}, \texttt{GPUs}, \texttt{tasks}, and \texttt{pools} are equal.}
\vspace{8pt}
\begin{threeparttable}
\scriptsize{
\setlength{\tabcolsep}{2pt}
\begin{tabular}{ccccccc}
\toprule
\toprule
 & \multicolumn{6}{c}{GPU}  \\
\midrule \\
\#CPU & 1 & 2 & 4 & 8 & 16 & 32  \\ 
\midrule
time (s) & 107640 &54780 & 27720 & 14100& 7500 &  3720 \\
time (m) &1794 & 913 & 462& 235& 125 & 
 62\\
 core-hours & 239 &  243 & 246 & 250 & 267 &  265\\
\bottomrule
\end{tabular}
}
\end{threeparttable}
\label{tab:6}
\end{table}

%% file: table7.tex
\begin{table}[ht]
\centering
\caption{Total time spent in the optimized \texttt{GPU} calculations. This includes also the time passed on the part of the code that are not GPU accelerated  or are not GPU optimized. The number of \texttt{CPUs}, \texttt{GPUs}, \texttt{tasks}, and \texttt{pools} are equal.}
\vspace{8pt}
\begin{threeparttable}
\scriptsize{
\setlength{\tabcolsep}{2pt}
\begin{tabular}{ccccccc}
\toprule
\toprule
& \multicolumn{6}{c}{optimized GPU} \\
\midrule
\#CPU & 1 & 2 & 4 & 8 & 16 & 32 \\ 
\midrule
time (s) &  16080 & 7860 & 4680 &  2153 & 1140 & 585\\
time (m) & 268 & 131 & 78 & 36 & 19 & 10 \\
core-hours &  36 &  35 & 42 & 38 &  41 & 43 \\
\bottomrule
\end{tabular}
}
\end{threeparttable}
\label{tab:7}
\end{table}